\documentclass[manuscript]{aastex}

\slugcomment{}

\shorttitle{The tidal filament of NGC 4660}
\shortauthors{Djorgovski et al.}

\begin{document}

\title{The tidal filament of NGC 4660}


\author{S. N. Kemp$^1$, C. Mart\'\i nez-Robles$^2$, R. A. M\'arquez-Lugo$^2$, D. Zepeda-Garc\'\i a$^2$, R. Franco-Hern\'andez$^1$, A. Nigoche-Netro$^1$, G. Ramos-Larios$^1$, S.G. Navarro$^1$, L.J. Corral$^1$ }
\affil{1. Instituto de Astronom\'\i a y Meteorolog\'\i a, Universidad de Guadalajara, Av. Vallarta 2602, Arcos Vallarta, 44130, Guadalajara, Mexico \\
2. CUCEI, Universidad de Guadalajara, Av. Revoluci\'on No. 1500 S.R., 44420 Guadalajara, Jalisco, Mexico}
\email{snk@astro.iam.udg.mx}







\begin{abstract}
NGG 4660, in the Virgo cluster, is a well-studied elliptical galaxy which has a strong disk component (D/T about 0.2-0.3). The central regions including the disk component have stellar populations with ages about 12-13 Gyr from SAURON studies. However we report the discovery of a long narrow tidal filament associated with the galaxy in deep co-added Schmidt plate images and deep CCD frames, implying that the galaxy has undergone a tidal interaction and merger within the last few Gyr. The relative narrowness of the filament implies a wet merger with at least one spiral galaxy involved, but the current state of the system has little evidence for this. However a 2-component photometric fit using GALFIT shows much bluer B-V colours for the disk component than for the elliptical component, which may represent a residual trace of enhanced star formation in the disk caused by the interaction 1-2 Gyr ago.  There are brighter concentrations within the filament which resemble Tidal Dwarf Galaxies, although they are at least 40 times fainter. These may represent faint, evolved versions of these galaxies.  A previously detected stripped satellite  galaxy south of the nucleus  is seen in our residual image and may imply that the filament is a tidal stream produced by perigalactic passages of this satellite.

\end{abstract}


\keywords{galaxies: structure, galaxies: photometry, galaxies: evolution,  galaxies: individual (NGC 4660),  galaxies: stellar content, galaxies: fundamental parameters}

\section{Introduction}
NGC 4660 is an E5 elliptical galaxy with a strong incrusted disk component, and has frequently been used as a `classic' example of this type of galaxy. As such it has appeared frequently in the literature. It is located in the Virgo cluster, and has been identified as a member of the nearest compact group satisfying the Hickson criterion \citep{mam89}, along with other Virgo galaxies like M59 and M60, but \citet{mam08} concluded from surface brightness fluctuation measurements that M59 and NGC 4660 form a pair $\sim 1 - 2$ Mpc closer to us than the other three galaxies of the potential compact group.   Hence we adopt a distance of 18 Mpc for this galaxy, giving a scale of 87 pc/arcsec and 5.2 kpc/arcmin.

Generally it has been assumed that the disk in NGC 4660 is primordial and that the galaxy has not undergone a significant interaction or merger, but here we report the discovery of a long, curved filament apparently emanating from NGC 4660, in the direction of M59, which implies that it has experienced some kind of major event with a dynamical timescale of a few $\times 10^{8}$ years.

The filament was discovered on a co-added array of scanned Kodak  Technical Pan films of the SE region of the Virgo cluster taken with the UK Schmidt Telescope \citep{kat98, kat01} which have previously been used to identify or confirm other filamentary structures, e.g. the filamentary `ellipse' of IC 3481 \citep{per09}, the filamentary connections between NGC 4410 A/B, C and D \citep{per08} etc.
Although such Schmidt plate/film studies have become almost obselete in modern astronomy, this discovery of the present, previously unsuspected, filament of NGC 4660 shows that they could still be put to productive uses.

Here we report the discovery of the filament using the Schmidt data, and describe subsequent multiband CCD observations  with the 2.1m telescope at the San Pedro M\'artir Observatory, which provide colour information for this galaxy and further imaging of the filament.  In Section 2 we provide a summary of  the previous work on the galaxy NGC 4660, while in Section 3 we describe the observational data and  their processing. Results of the photometry of NGC 4660 and of the filament are given in Section 4, and in Section 5  we discuss the age and the formation of the filament and give general conclusions. 

\section {Previous work  on NGC 4660}

  NGC 4660 (VCC 2000) has been studied in various ways, and has formed part of many samples of early-type galaxies over the years, on account of its proximity to us in the Virgo cluster and its incrusted disk component. It is located at RA = 12:44:32.0 Dec. = +11:11:26 (J2000), has a recessional velocity of 1083 km/s, and redshift of 0.003612 (NASA/IPAC Extragalactic Database).  Although its classification in the SIMBAD database is E5, the Virgo Cluster Catalogue \citep{bin85} gives a classification of E3/S0--1(3), which emphasises its transitionary nature.

 There have been a few previous photometric and structural studies of NGC 4660, though three of them  have used the same observational dataset. We  compare our photometric results with theirs in Section 4. \citet{benet88} carried out CCD photometry and isophote analysis on NGC 4660 in the V, R and I bands. They found an effective radius $r_{e}$  between 9--9.8 arcsec and an isophote twist of $5^{\circ}$. Peak ellipticity was $\sim 0.5$ while $a_{4}/a$  (used as the 4th cosine coefficient in this case) rises to 0.04 at 30 arcsec. 

\citet{rix90} performed photometric modelling of galaxies with 2 components (spheroidal $r^{1/4}$ law bulge plus exponential disk). They applied their techniques to \citet{benet88}'s  data on NGC 4660 and found they achieved better fits with a significant disk component included ($L_{disk}$ $\sim 50$\% $L_{bulge}$). \citet{ben88} described NGC 4660 as a rapidly-rotating elliptical while \citet{rix90} conclude that it is the disk rotation which is the major contributor to the detected velocity. 

\citet{sco95}, again using the same photometric data of \citet{benet88} for NGC 4660, find disk signatures in the ellipticity and $a_{4}/a$ profiles  (again used as the 4th cosine coefficient).  The ellipticity is $\sim 0.2$ at the centre and peaks at 0.5 at 10--30 arcsec, while $a_{4}/a$ is $< 0.01$ in the centre, rising to 0.02--0.04 between 10--40 arcsec. The velocity dispersion $\sigma$ is $\sim 200$ km s$^{-1}$  at $< 5$ arcsec  (0.43 kpc) and 100 km s$^{-1}$ at 30 arcsec (2.6 kpc), while the disk rotation velocity is 180 km s$^{-1}$ at 10--25 arcsec (0.87--2.17 kpc) and $\sim 100$ km s$^{-1}$ at 0--10 arcsec  ( 0--0.87 kpc). They find evidence of an isophotal twist, with PAs in the range $90^{\circ}$ to $100^{\circ}$, between 10--20 arcsec (0.87--1.74 kpc), taking this as evidence of a disk with inclined bar extending to 12 arcsec (1.04 kpc) from the centre. Its surface brightness profile is similar to early-type barred galaxies \citep{com93}. For the disk, $v/{\sigma}$ is very high, $\sim 3.3$, and D/B within the bar is 0.5. \citet{sco95} suggest that the disk may be unstable to bar formation, and the bar may have a long evolution time.

The dataset of \citet{sco95} comprises mainly disky elliptical galaxies i.e. E galaxies with `pointed' isophotes (of which, NGC 4660 has the highest D/B ratio, 0.28). These galaxies are found to lie on the same correlation between central surface brightness and disk scale length defined by lenticular and spiral galaxies. This implies that disky E's are not inclined S0's but form `transition' objects between E's and S0's. The disk profile is not generally exponential. In NGC 4660 the disk and bulge rotate in the same direction and have similar surface brightness profiles, suggesting that  they formed at about the same time out of the same material.

 This galaxy was observed by \citet{fer06} with HST/ACS as part of a sample of 100 early-type galaxies in Virgo. These data  deal with the central regions of the galaxy at higher spatial resolution, and so the present data  are complementary to this dataset. Their image shows the presence of a possible stripped satellite galaxy, 2.5 arcsec across and 4.5 arcsec south of the centre, described as blue, `resembling a nucleus with two spiral arms'. They fit the luminosity profile  of NGC 4660 with one component with S\'ersic indices of 4.0 in the $g$ band and $4.5$ in the $z$ band and an average $g-z$ colour of 1.51. Structural parameters such as ellipticity, PA, $a_{3}, a_{4}, b_{3}, b_{4}$ are reported, giving results quite similar to previous references.  $b_{4}$ is used here as the 4th cosine coeffcient  and provides a clear separation of the boxy bulge and the disk. The $g-z$ colour profile shows a blueward tendency from 0-3 -- 20 arcsec then rises slightly to 50 arcsec.

NGC 4660 forms part of the SAURON sample, a survey of 72 early-type galaxies using the integral-field spectrograph SAURON at the William Herschel Telescope. Stellar population studies of this sample give us ages and metallicities of the dominant stellar populations in central regions of this galaxy \citep{bac01, zee02}. \citet{kun10} carried out a stellar population analysis of absorption line strength maps of 48 of these galaxies including NGC 4660. They find only old stellar populations in this galaxy, $12.2 ^{+ 1.2} _{-0.6}$ Gyr at $R_{e}/8$   (1.4 arcsec, 1.3 kpc) and $13.4 ^{+1.3} _{-1.2}$ Gyr at $R_{e}$  (11.5 arcsec, 1.0 kpc). The metallicity [Fe/H] varies from $0.15 \pm 0.02$ at $R_{e}/8$ to  $0.11 \pm 0.02$ at $R_{e}$, while the $\alpha$-element composition varies slightly, [$\alpha$/Fe] is $0.24 \pm 0.04$  at $R_{e}/8$ and $0.29 \pm 0.05$ at $R_{e}$. Although the [$\alpha$/Fe] variation is within the errors, it is typical of the other galaxies in the sample to have depressed [$\alpha$/Fe] values in the centre corresponding to higher metallicities. The velocity dispersion $\sigma$ is 221 km s$^{-1}$ at $R_{e}/8$ and 181 km s$^{-1}$ at $R_{e}$. All galaxies which are fast rotators in the sample and which have flattened components with disk-like kinematics are found to have different stellar populations in these flattened components. Those with young populations  ($ < 3$ Gyr) frequently have circumnuclear disks and rings which are still forming stars. Meanwhile NGC 4660 belongs to the other extreme, in which the structure with disk-like kinematics has slightly older ages and lower [${\alpha}$/Fe] ratios in which the `secondary' star-formation event which formed the disk-like structure is presumably nearly as old as the `elliptical'  component of the galaxy.  We compare these results with populations indicated by our photometric colours in Section 5.

Another interesting line of work carried out concerning NGC 4660 is its possible membership of a Compact Group (CG).    \citet{mam89}  considered it, alongside M59, M60, NGC 4638 and NGC 4647, as a possible CG involving members of the Virgo cluster, as it was found to satisfy Hickson (1982)'s criteria according to new magnitude measurements. However, \citet{mam08} concluded from surface brightness fluctuation measurements  of individual early-type galaxies that NGC 4660 may form a pair with M59, at a distance of $\sim 2$ Mpc closer to us than most of the rest of the Virgo cluster. Although the various statistical calibrations differ, M59 and NGC 4660 as a pair are at least 440 kpc closer to us than M60 and NGC 4638. This line-of-sight depth would be too great for a CG \citep{mam08}.   Our new data do not provide anything new on this interesting possibility.

In the following section we consider the new observational data and  their processing.

\section{Observations and Data Reduction}

\subsection{The Schmidt films}

The use of deep, wide-field imaging to detect faint material in groups and clusters is well-established, e.g. \citet{kem93}. \citet{kat98, kat01} produced a stack of 13 digitally co-added Kodak Technical Pan films obtained with the UK Schmidt telescope, using the `OR'  filter (equivalent to the Cousins R band). The films covered the SE part of the Virgo cluster, with the aim of detecting and mapping the faint material in galaxy haloes and filamentary structures in this cluster. Faint material is detected down to 28 $R$ mag arcsec$^{-2}$.

Figure 1 is a section of these data containing NGC 4660. The curved filament is clearly seen to the NW of the galaxy.  Our CCD fields are indicated and an arrow gives the direction of M59.

\subsection{Optical CCD photometry and data reduction}

We carried out direct imaging of a field containing NGC 4660 (at its SE corner, to include part of the area occupied by the filament) on 2005 March 30 with the 2.1m telescope of the Observatorio Astron\'omico Nacional, San Pedro M\'artir, Baja California, Mexico. The SITe1 CCD camera, with  Johnson BVRI filters was used. We obtained total combined exposure times of 900, 600, 450, 300s in BVRI respectively  in photometric conditions.  The spatial scale for these data was 0.31 arcsec pixel$^{-1}$ and the CCD field of view was $5.3 \times 5.3$ arcmin$^2$, with a seeing of $\sim 2$ arcsec.
6 images of 600s in R were also taken of a field to the W which is centred on the brightest parts of the filament (as seen in the Schmidt data).

The standard procedures of bias-subtraction, dark-correction and flat-fielding were carried out using IRAF. \footnote{IRAF is distributed by the National Optical Astronomy Observatory, which is operated by the Association of Universities for Research in Astronomy (AURA) under a cooperative agreement with the National Science Foundation.} 

For the broad-band observations, we observed Landolt standard fields containing Palomar Green stars  \citep{lan92} at intervals during the night at a range of airmasses. Estimates of the extinction coefficients were obtained by plotting instrumental magnitudes against airmass for multiple observations of the same field of standard stars. These values of the extinction coefficients were used to obtain photometric zero points and colour terms. The photometric transformation equations obtained were
\begin{equation}
B = b + 24.24  - 0.24X + 0.13(B-V), 
\end{equation}
\begin{equation}
V = v + 24.94 - 0.15X  + 0.02(B-V), 
\end{equation}
\begin{equation}
R = r + 24.67 - 0.01X  + 0.05(B -V), 
\end{equation}
\begin{equation}
I =  i  + 24.17 -0.04X + 0.05(B-V),  
\end{equation}
where $X$ is the effective airmass of the exposure.

The errors associated with our data have two main contributions, a) the systematic error in obtaining the calibration equation, b) the error in determining the true sky value.
The first is obtained by analysing the residuals in the fits used to obtain the parameters of the calibration equation. The second is estimated by measuring the sky level in several different areas of the image (free from bright objects) and calculating the $\sigma$  for these independent measurements of the sky value. Error a) is the most significant in brighter regions whilst b) is most significant for fainter regions.  Errors in colours are calculated by adding in quadrature errors in individual bands.

We note that for the 2.1m telescope data we have used Landolt standards defined in the Johnson-Kron-Cousins system while the filters used  for the observations are Johnson filters, and in R and I there are significant differences  between the two filter systems. \citet{per08} reported no significant differences between the surface brightnesses obtained over a range of galaxy colours, using previous observations of galaxies in both systems.

 The typical seeing  FWHM in each filter was measured and the images smoothed with a gaussian filter until the image in each filter had the same FWHM. The calibration equations were applied to the counts in the images to obtain colour maps by direct subtraction of images.

We performed surface photometry on NGC 4660 using the final images in counts with the stars masked out. Radial colour profiles were created by subtraction of individual calibrated surface brightness profiles, with errors calculated as above. Ellipse-fitting was carried out using the program {\it ellipse} in the  isophote package in IRAF, with an outer isophote of  25.8 $B$ mag arcsec$^{-2}$ in B (4\% of sky level).

The best-fit isophotal models generated can be subtracted from the final reduced image (in counts) to produce residual images, in which non-axisymmetrical structures previously hidden in the brighter parts of the galaxy can be revealed. 

We also used GALFIT \citep{pen02} which is an algorithm that can analyse the  light distribution profiles of astronomical objects in digital images, using analytic functions, using single-component models to describe the overall morphology or using a combination of different structural components  such as bulges, disks, bars, spiral arms, to model more precisely the structure of particular galaxies.  Each component can be described by scale lengths, total magnitudes, ellipticity, position angle, S\'ersic index etc. We used both a one-component S\'ersic model  and  a two-component model (S\'ersic and exponential disk) to fit the images of NGC 4660 in all photometric bands, obtaining S\'ersic indices, bulge-to-disk luminosity ratios (B/D), effective radii, scale lengths etc.  A star taken from the same field  was selected as the input PSF image in GALFIT, which is included in the modelling fits to the image.  This PSF image is 101 x 101 pixels with its peak flux at the 51st pixel in X and Y.

\section{Results}
\subsection{NGC 4660}

An image of the CCD field in V containing NGC 4660 is presented in Figure  2a. 
In order to show details across the entire range of brightness and the brightest core regions, the image 
was represented in logarithmic scale and processed through layer masks. As mentioned in Section 3.2 the galaxy appears in the SE corner, to allow the inclusion of part of the area occupied by the filament in the field, though this part of the filament was not detected even at high contrast. The `pointy' isophotes indicating the presence of a disk component are quite obvious. There is a small diffuse object about 1 arcmin north-east of the centre of NGC 4660,  VCC 2002 (LEDA 49295) which is classified as a dE \citep{bin85}.  The residual image in V is shown in Fig. 2b. A characteristic pattern due to the disk component is evident, with excesses of light (bright areas) along the major and minor axes and minima (dark areas) at $\sim 45^{\circ}$ to these.  The stripped satellite galaxy observed by \citet{fer06} can be seen at about 5 arcsec South of the centre of NGC 4660.

The surface brightness profiles in the 4 filters are shown in Figure  3a.  The outer ellipse has a semi-major axis of approximately 100 arcsec. The centre of this galaxy is particularly bright, reaching a central surface brightness of 14.7 $I$ mag arcsec$^{-2}$, 17.1 $B$ mag arcsec$^{-2}$. Fig.  3b shows the surface brightness profiles plotted against $a^{1/4}$, showing that this galaxy does not obey the de Vaucouleurs $r^{1/4}$ law, while at the same time, does not show huge deviations from this law.  The profiles in each of the 4 filters show the same broad tendencies.
 
Figures  4a and 4b show the $B-V$ and $B-R$ colour maps, respectively. The most distinctive feature is the nuclear and central region, including much of the disk, which displays slightly redder colours to a radius of about 20 arcsec. NNE of the nucleus, both $B-V$ and $B-R$ display a slightly yellower zone.  The `halo', i.e. the main body of the elliptical, is much noisier but is bluer than the centre. In $B-R$ the eastern half of the disk appears slightly less red than the western, by $\approx 0.10 - 0.15$ mag. The small galaxy VCC 2002, 1 arcmin NE of the centre, also can be seen in these colour maps.  There is no clear evidence of the  feature at 4.5 arcsec south of the centre as in the HST/ACS data of \citet{fer06},
though there does seem to be a slight and abrupt change in colour,  becoming redder going from east to west, at its position South of the centre of NGC 4660.  There does seem to be some subtle colour changes over the disk area, which may support the  `tidal stream' hypothesis (see Section 5). 

Figure  5 shows the radial colour profiles of NGC 4660 in $B-V$, $B-R$ and $B-I$. The error bars include both systematic and sky noise contributions. In the central regions the sky noise is insignificant relative to the galaxy counts, so changes in colour of order 0.05 mag are significant in  these regions. $B-V$ is at about 0.90-0.95 in the disk region, rising slowly to $\sim 1.15$ at $a \sim 70-80$ arcsec then declining again. In $B-R$ the disk colour is $\approx 1.7$, rising to 1.90-2.00 at $\sim 70-80$ arcsec, then declining further out. In $B-I$ the regions occupied by the disk have a colour of $\approx 2.3$, and this rises steadily to about 2.8 at 65--80 arcsec. In general there is  evidence for a slight but significant redward  gradient in the fitted isophotes from about 35 to 80 arcsec, while the region occupied by the disk has a flatter colour profile.  The colour maps however do appear to show that the central plane of the inner disk component may be $\sim 0.1$ mag redder than neighbouring parts of the galaxy. However the galaxy light is not dominated by the disk even at small radii and the overall colours of the isophotes (rounder than the disk component) may be less red than implied by the disk colours. The $g -z$ colour profile of \citet{fer06} is dominated by a blueward tendency between 0.3 and 20 arcsec but  is redward to
80 arcsec and so is broadly in agreement with our profiles.

For all the structural parameters investigated, the $V$ band data  are used, although no significant variation was seen between this and other filters. The radial variations of the structural parameters $a_{4}/a$ and $b_{4}/a$ are plotted in Figures  6a and b respectively.  We use $b_4/a$ as the 4th cosine coefficent.  $a_{4}/a$ is predominantly slighty positive a (though $< 0.01$) between 10 and 30 arcsec where the disk dominates, and predominantly slightly negative  between 30--45 arcsec. 
The value of $b_{4}/a$ rises steadily from 0.01 at $\sim 10$ arcsec to 0.05 at $\sim 30$ arcsec, indicating clearly disky isophotes in the whole radial range occupied by the disk. Beyond $\sim 45$ arsec the values of $b_{4}/a$ are negative (boxy),  implying a disk embedded in a boxy halo. The same tendencies with radius are shown in the HST/ACS data of \citet{fer06} between 5--30 arcsec  (this paper also uses $b_4/a$ as the 4th cosine coefficient.)

The radial ellipticity profile of NGC 4660 is shown in Figure  7a.  The Position Angle profile is shown in Figure  7b.  Both the ellipticity and PA profiles are similar to other results in the literature   including \citet{fer06}, the twist at 5--10 arcsec being taken as evidence for a disk with inclined bar \citep{sco95}. The bulge/halo therefore has the same position angle as the disk component.

  Results from the GALFIT program show that the bulge component and the galaxy as a whole has a $(B-V) \approx 1.0$ while the disk is significantly bluer with $(B-V) \approx 0.7$.  The D/B ratio is 0.09 for B, and 0.06 for V  i.e. confirming that the disk is somewhat bluer than the bulge component. These are rather lower D/B ratios than previously reported.  S\'ersic indices for the bulge component are $\approx 6$. The disk scale length of around 11 arcsec is similar to the galaxy effective radius of previous authors, e.g.\ \citet{kun10}, while the bulge component effective radius is twice as large, maybe due to our relatively deeper exposures detecting more light associated with the filament around the galaxy. This may also explain the lower D/B ratios.  For a summary of photometric parameters measured, see Table 1.

\begin{table}
\caption{ Photometric parameters measured for the whole galaxy, bulge and disk component in the $V$ filter}
\begin{tabular}{crrr}
Parameter &  Total & Bulge & Disk \\
Magnitude  & 10.9 & 11.0 & 14.1 \\
$r_e$ (arcsec) & 22  & 21 &  \\
$r_s$ (arcsec) &  &  & 11\\
$n$     & 5.8    &  6.2  & 1  \\
Ellipticity & 0.3 & 0.3 & 0.6 \\
PA (degrees) & 91 & 91 & 91 \\
color $B-V$ & 1.0 & 1.0 & 0.7 \\
D/B &  0.06 & & \\
\end{tabular}
\end{table}

\subsection{The filament}
The brightest parts of the filament as seen in the Schmidt image (Fig. 1) are detected in the CCD field (Figure  8)  as an area of diffuse emission in the north-centre of the field  (running NNE-SSW)
and just S of centre between two stars. There is also an area of emission in the SE of the field corresponding to emission from the NW part of the halo of NGC 4660. The northern detection of the filament  has two peaks of surface brightness of approximately 26.6 $R$ mag arcsec$^{-2}$ (RA = 12:44:18.5 Dec. = +11:15:27 (J2000)) and 27.4 $R$ mag arcsec$^{-2}$ (RA = 12:44:20.2 Dec. = +11:16:04, corresponding to only about 0.4\% and 0.2\% respectively of the sky brightness in $R$. These diffuse features may correspond to Tidal Dwarf galaxies (TDGs).  Figure 9 shows the Schmidt field containing  NGC 4660 and the filament without any lines marked on the image, giving a clearer unobstructed view of the filament, in which the brightest areas in the CCD filament field can be seen more clearly (with help from Fig.\ 1).

The northern object (TDG-1) has an apparent $R$ mag of approximately 20.9 while that of TDG-2  is 21.5, corresponding to absolute magnitudes of --10.4 and --9.8 respectively. The scale lengths are of the order of a few pixels (100-200 pc).
There are further candidate objects to the South marked as `TDGs?'.

The straightened-out length of the filament is about 12 arcmin, corresponding to about 60 kpc at a distance of 18 Mpc to NGC 4660. Moving at a typical velocity for material in filaments of 100 km s$^{-1}$ (for example \citet{hib95}), stellar material at the tip of the filament would have taken $\sim 6 \times 10^{8}$ years to reach this position.

\section{Discussion and Conclusions}

Tidal tails and other fine low surface brightness features around galaxies can only be formed in gravitational interactions between galaxies (usually major mergers with proportion of masses less than 4) over a timescale of a few $\times 10^8$ years \citep{tom72}. Therefore the detection of the filament near to NGC 4660 appears to indicate a past gravitational interaction for this galaxy, for which there was no previous evidence. Properties of tidal tails can be used to date the merger event which formed them.     Tidal tails may be long-lived after a gravitational encounter.

NGC 4660 may have had an encounter, or a `dry' interaction (not merger), with another early-type galaxy. The nearest candidate galaxy for such an interaction with NGC 4660 is LEDA 42878 (VCC 1991) which can be seen at about 6 arcmin { West-SouthWest of NGC 4660  (seen in Figs.\ 1 and 9), at 12:44:09.2 +11:10:32 (J2000),  is classified as a dE in the SIMBAD database and identified as a nucleated dwarf galaxy by \citet{bin85}. This has a radial velocity of 1681 km s$^{-1}$ \citep{san11}, approximately 600 km s$^{-1}$ more than that of NGC 4660, which would make any significant interaction between them unlikely. There is also the dE 1 arcmin NE of the centre of NGC 4660, VCC 2002, but this appears to be too small to produce such a long filament in an interaction with NGC 4660.

Such prominent tidal tails are more likely to be formed in `wet mergers' (in which at least one progenitor is a gas-rich spiral). There are no H I observations which could demonstrate  whether the
filament observed near NGC 4660 is gas-rich. The presence of only one tail may indicate that only one of the progenitors of  NGC 4660 was a gas-rich spiral, 
The period of time for which a tail may be observed is important in studies of galactic history and evolution, especially in the case of NGC 4660 where it was the only direct evidence of the
galaxy having experienced a tidal interaction.

Tails are formed on the dynamical timescale  of $\sim 10^{8}$ years and the existence of a bright tail does not imply that the interaction which formed it was recent, bright tails may survive for a few Gyr. 
The galaxy evolution numerical simulations of \citet{pei10} show early-type/spiral mergers producing peaks in Star Formation Rate (SFR) for $\sim 2$ Gyr after closest approach, while shells are
seen at $\sim 1.5$ Gyr after closest approach. 
\citet{con09} predicts maximum merger timescales of $1.1 \pm 0.3$ Gyr, without taking into account the detailed history of the fallback of tidal features. 

\citet{hib95} modelled the spatial morphology and velocity structure of the famous merger product
NGC 7252, suggesting that the merger occurred 0.6 Gyr ago and 80\% of the mass
would fall back to the merger remnant in 2.5 Gyr. So detection of tidal debris would get difficult after about 3 Gyr, while if the merger  were older there would be time for more minor mergers to
occur, disrupting the filament \citep{duc11}. Most of the tail material would and does remain bound -- there are currently velocity reversals along the tails  indicating material which has already reached the turnaround point in its orbit and has started falling to smaller radii. 20\% of current tail particles will not fall back to $< 5 R_{e}$ in a Hubble time. A filament as  relatively narrow  as the one in NGC 4660 would require at least one of the parent galaxies to be a spiral galaxy, though the other may have been an early-type dynamically hot galaxy, which would only have produced plume-like debris which may disappear more rapidly. Alternatively, if the other galaxy was indeed another spiral galaxy,  the other tail has evaporated or was
destroyed in a subsequent minor merger interaction with the general gravitational potential well of the cluster/group environment. So in Virgo one may expect tails
to survive for maybe less than 2 Gyr.

The presence of a spiral galaxy would make this a gas-rich (`wet') merger. The two peaks in surface brightness in the  North of the CCD field (Figure  8) may correspond to Tidal Dwarf Galaxies (TDGs) which are gravitationally bound small systems of stars and gas formed in major mergers (mass ratio $> 1:4$), see \citet{mir92}. TDGs are generally thought to form from gas clouds pulled out of galaxies during mergers and so do not represent  previously existing stellar systems. Here we only have data in red filters for these TDGs, which are often found to have a bluish colour \citep{duc11}. Studies of the colours and spectra of the candidate TDGs in NGC 4660 would prove interesting, but are made difficult by their faintness.
However the candidate old tidal dwarfs studied by \citet{duc14} have typical sizes of 0.8--2.3 kpc and absolute magnitudes corresponding to --13.5 to --17.5 in $R$, while their central surface brightnesses are 23.5--26.5 $R$ mag arcsec$^{-2}$. Younger dwarfs in the same article have scale lengths of 2--6 kpc,  absolute $R$ magnitudes of --14 to --17.5, and central surface brightnesses ranging from 20.5 to 25 $R$ mag arcsec$^{-2}$. While our candidate TDGs in NGC 4660 are at the low end of the surface brightness range for old dwarfs they are much smaller objects, with luminosities at least 40 times less than previously identified TDGs, and only a few pixels in size, so they inevitably have small scale lengths given the detection threshold. The implied masses will also roughly be 2 orders of magnitude less, so while the objects in \citet{duc14} are of the order  of $10^8 M_{\odot}$, the present objects may be only about $10^6 M_{\odot}$ in mass.
 \citet{duc11} consider that the TDGs in the eastern tidal tail of NGC 5557 may be at least 2 Gyr old and they are still 1.3 mag bluer than the rest of this tail, while \citet{duc14}  give a spectroscopic age of 4 Gyr for one of these TDGs. 
If the filament in NGC 4660 is several Gyr old then these could represent faint, evolved, even older TDGs, however they are so much smaller that they seem to represent extreme lower-luminosity cases of the class of TDGs or a different class of objects. Evidence for long lived TDGs,  or detection of many fainter but similar objects, would be significant  in terms of studies of numbers of dwarf satellite galaxies.

The `stripped satellite' detected by \citet{fer06} may provide an alternative origin for the filament, formed by stars tidally stripped from this satellite galaxy during the closest parts of its orbit to NGC 4660, making the filament a tidal stream, comparable to the tidal stream associated with the Sagittarius dwarf galaxy satellite of the Milky Way \citep{iba94}. This would account for the presence of only one detectable filament. \citet{sta15} note that detection of extragalactic tidal streams from the ground is complicated by their low surface brightnesses, below 28 mag arcsec$^{-2}$ and this seems just about compatible with the surface brightness of the filament of NGC 4660 away from the possible TDGs. The colour maps of  Figure 4a and 4b do imply some complicated structure in the inner disk area, with gradients from east to west, on the South side of the disk.  Studies of tidal streams are now commonly made with the 3.6$\mu$m and 4.5$\mu$m filters of the {\it Spitzer Space Telescope} on its warm mission \citep{sta15}. It would be interesting to map the area around NGC 4660 in these filters. While the present filament is relatively bright, this tidal stream hypothesis does provide a possible explanation for the origin of the filament.

The ages of stellar populations in the galaxy given by the SAURON data \citep{kun10} are 12--14 Gyr, with metallicities of 1.3--1.4 times solar (with marginal evidence for lower central metallicities) though this only covers the galaxy interior to and at $R_e$ (11.5 arcseconds). The $(B-V)$ colour profile varies from $\sim 0.9$ in the disk region to $\sim 1.15$ in the exterior regions, which is a bigger colour change than implied by the marginal metallicity gradient in  the SAURON data (according to the Tables in \citet{bre94} the change in metallicity will only raise the $B-V$ colour by a few hundredths of a magnitude).
However, the GALFIT results do show a rather bluer colour for the disk ($B-V  \approx 0.7$ compared with $\approx 1.0$ for the much larger and brighter S\'ersic component), and the disk is progressively less prominent in the 2-component fit from B to I. The models of \citet{bre94} give $B-V$ colours of 1.03--1.13 for a 12.5 Gyr population with metallicity 1.0--2.5 times solar, slightly redder  than the S\'ersic component here but compatible with the colours in the inner colour profile. Meanwhile the disk component colours can be reproduced with models of age 1-2 Gyr for solar metallicity and 1 Gyr for 2.5 times solar metallicity.

So this may imply relatively young stellar populations in the disk, masked by the fact that the elliptical `bulge' component dominates at all radii, and because the SAURON data use circular apertures, and now revealed by the ability of GALFIT to separate the light from the two components. This may therefore be evidence for star formation produced by a merger 1-2 Gyr ago and now  detectable through a bluer colour in the  stellar population of the disk component separated by GALFIT.

 In conclusion, we have detected a tidal filament in the vicinity of NGC 4660 using the enhanced Schmidt film material, and detected brightness peaks of the filament in subsequent deep CCD data. This is not a galaxy expected to have undergone a tidal interaction/merger recently, and the SAURON data indicate only old populations, and a disk coeval with, or only slightly younger than,  the elliptical component of the galaxy. However the two-component fit shows that the disk component has bluer colours, indicating that the disk may have been formed in a merger 1-2 Gyr ago, or that this interaction caused enhanced star formation in a pre-existing disk. There are brighter concentrations within the filament which resemble Tidal Dwarf Galaxies, although  they are at least 40 times fainter. These may represent faint, evolved TDGs or may be another class of object. A previously detected stripped satellite south of the nucleus it detected in our residual image and may imply that the filament is really a tidal stream produced by perigalactic passages of this satellite. It would be interesting to carry out deep H I observations in the region of the filament, and its brightness peaks, to determine its gas content, and  to obtain {\it Spitzer} images of the filament, mapping the old stellar popualtions to fainter levels.

\acknowledgments
We thank CONACyT for providing financial help for the project ``Stellar populations in early-type galaxies''. We thank the staff of the UKST for obtaining the Schmidt films between 1991 and 1994,
the APM group for scanning the films, and Mike Irwin for helpful advice at that time. The co-addition of the APM scans of the Schmidt films was carried out at the Manchester STARLINK node by Dave Berry and Thanassis Katsiyannis. We thank John Meaburn for having the idea of stacking Schmidt plates. We also thank all the staff in San Pedro M\'artir for all their technical support,  and Violeta Guzman and Nely Cerda  for their help with the observations. Thanks to Andr\'es Rodr\'\i guez for computer management and software help. We thank the anonymous referee for many thorough and helpful suggestions, leading to a great improvement in this paper.

\begin{figure}
\epsscale{1}
\plotone{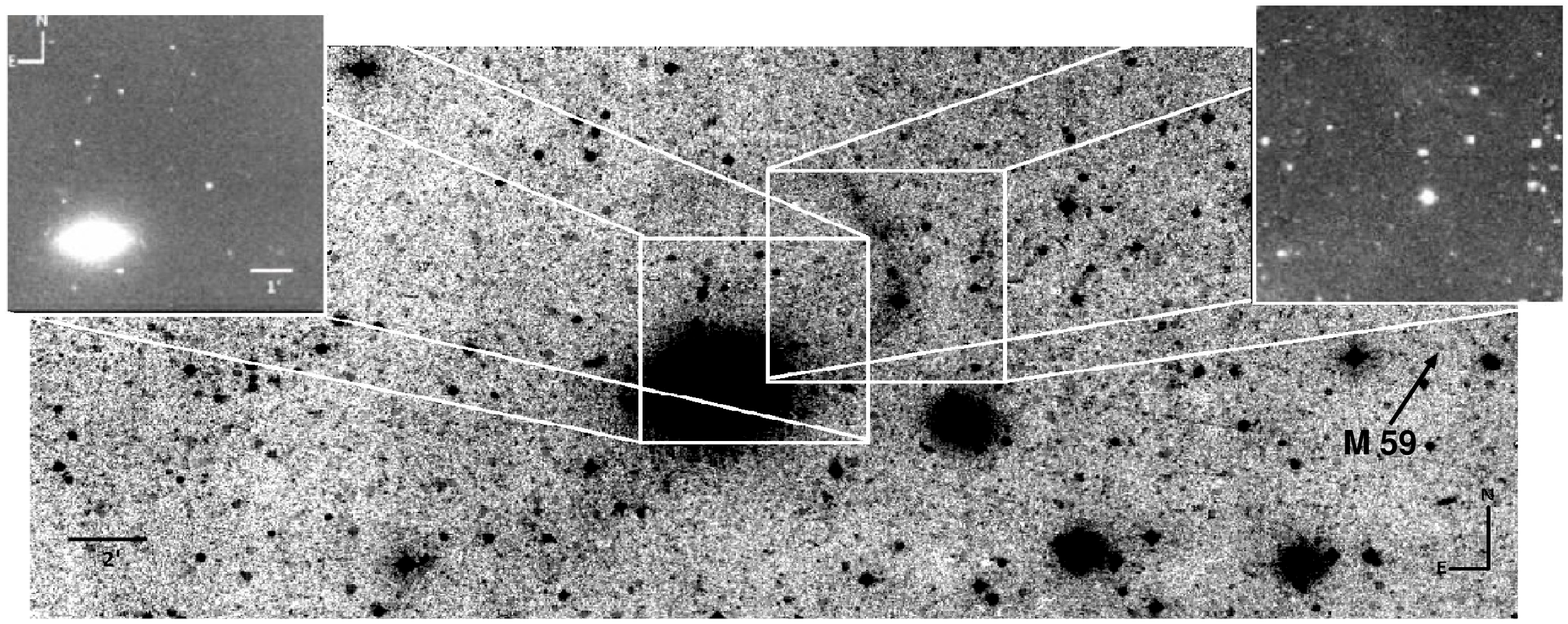}
\caption{Schmidt field of $30 \times 15$ arcmin$^{2}$ showing NGC 4660 with the two CCD fields, of $5.3 \times 5.3$ arcmin$^{2}$, superimposed.  North is to the top and east is to the left in each image. The direction of M59 (outside the field) is indicated by an arrow.}
\end{figure}

\begin{figure}
\epsscale{.55}
\plotone{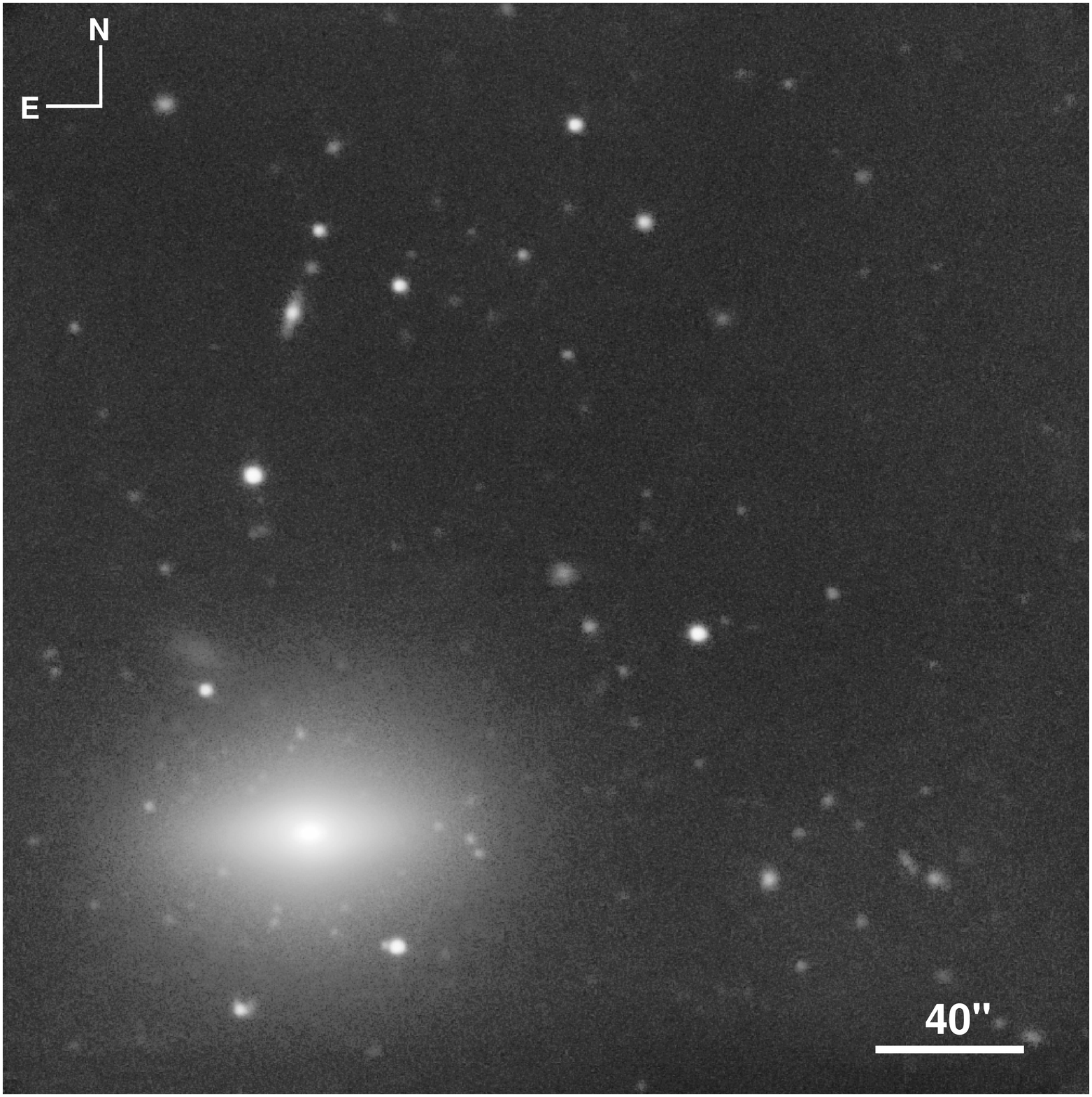}
\plotone{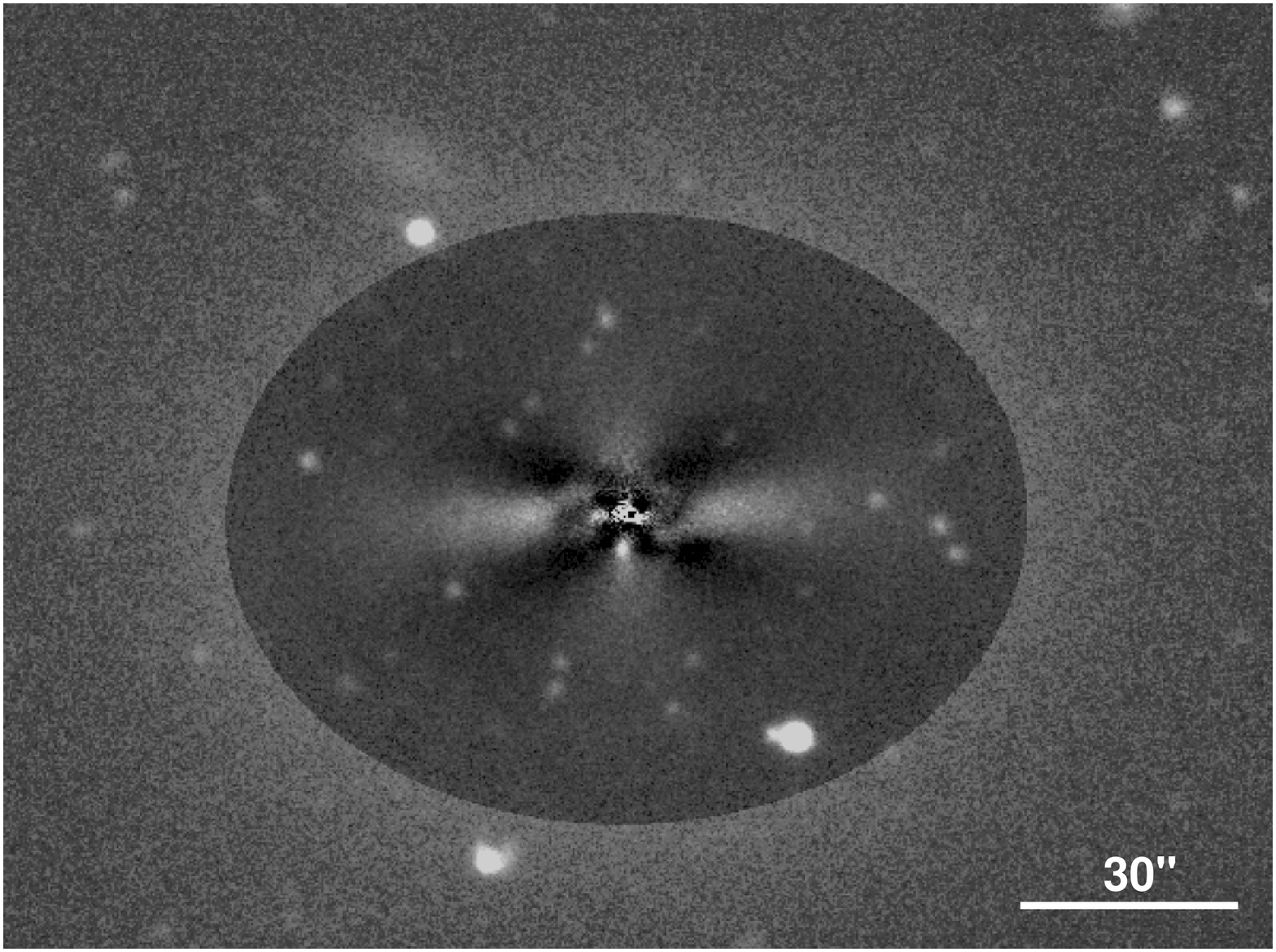}
\caption{ a) CCD field in V band containing NGC 4660, in the SE corner. Taken at the 2.1m telescope of the Observatorio Astron\'omico Nacional, Baja California, M\'exico. The field is 5.3 x 5.3 arcmin$^{2}$. North is to the top and east to the left. In order to show details across the entire range of brightness and the brightest core regions, the image 
was represented in logarithmic scale and processed through layer masks.  b) Residual image in V of NGC 4660 from the CCD data. The bright cross pattern in the centre of the galaxy indicates the presence of disky isophotes. The bright spot 5 arcmin south of the centre may be the `stripped satellite galaxy' detected by \citet{fer06}.}
\end{figure}

\begin{figure}
\epsscale{1}
\plotone{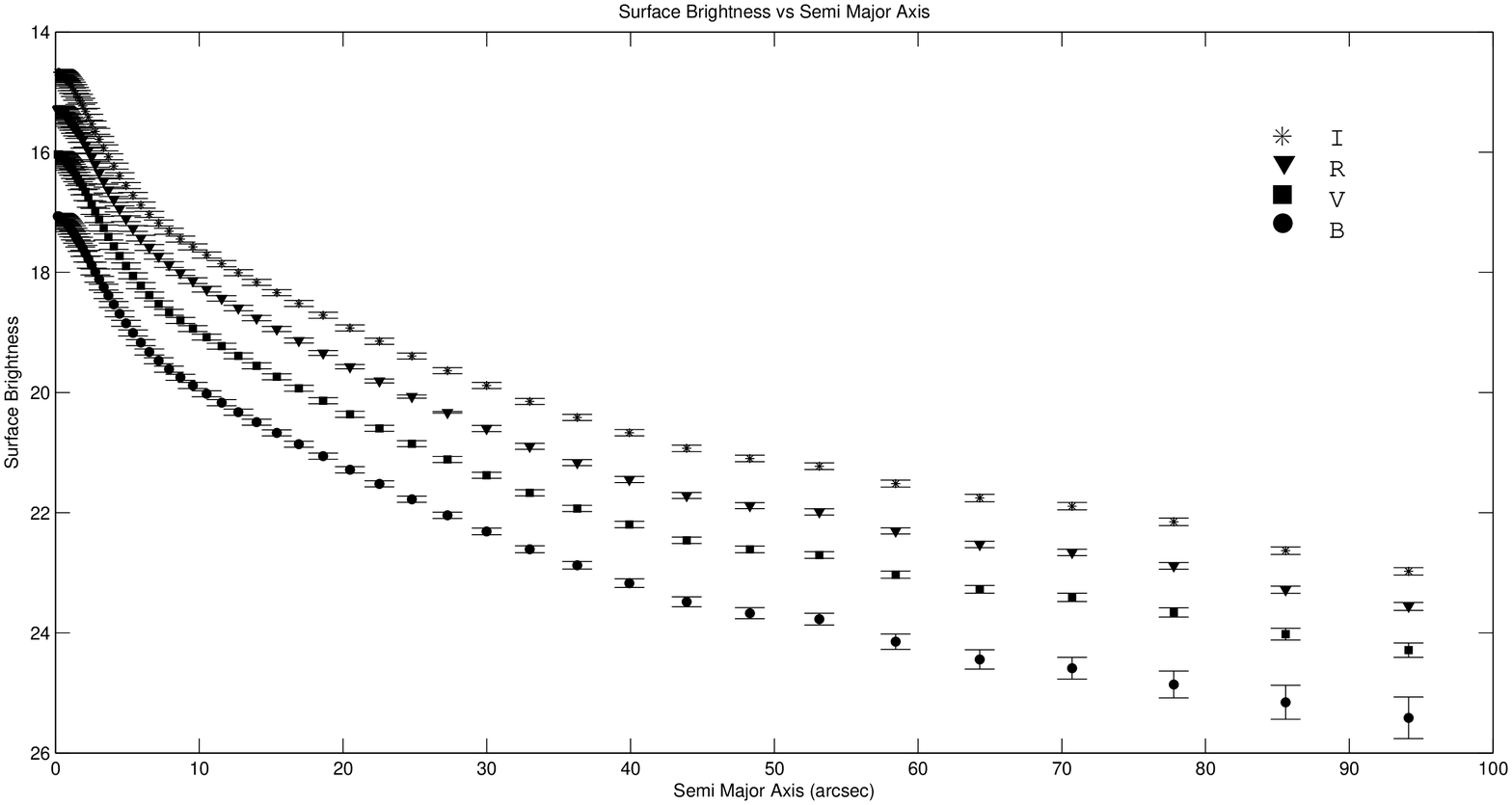}
\plotone{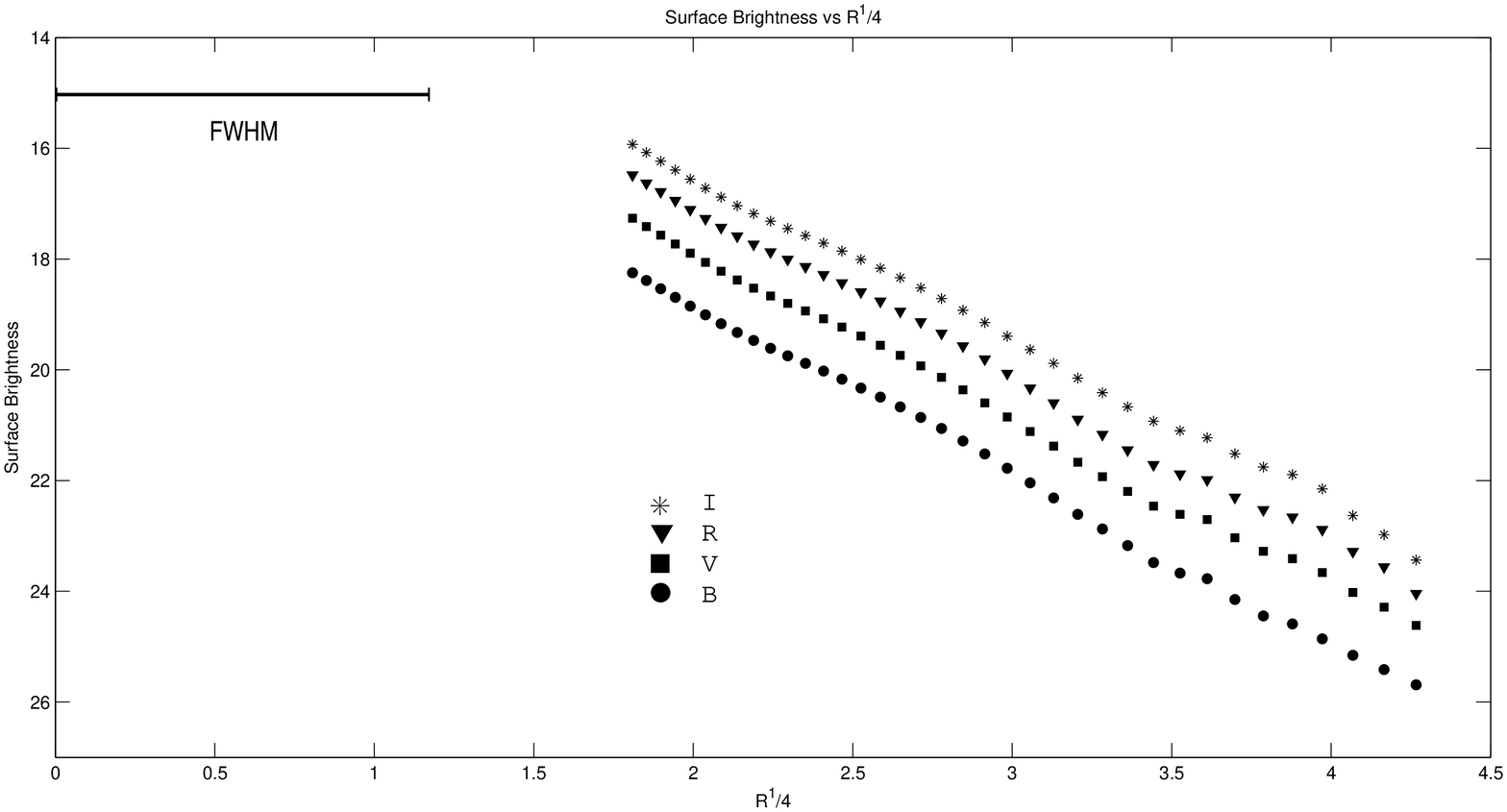}
\caption{a) Radial surface brightness profiles of NGC 4660 in B, V, R and I filters, from the CCD data. Error bars are displayed, calculated as explained in the text  b) $r^{1/4}$ surface brightness profiles of NGC 4660 in B, V, R and I filters from the CCD data. The seeing FWHM of the data is indicated by a a horizontal line.}
\end{figure}

\begin{figure}
\epsscale{.70}
\plotone{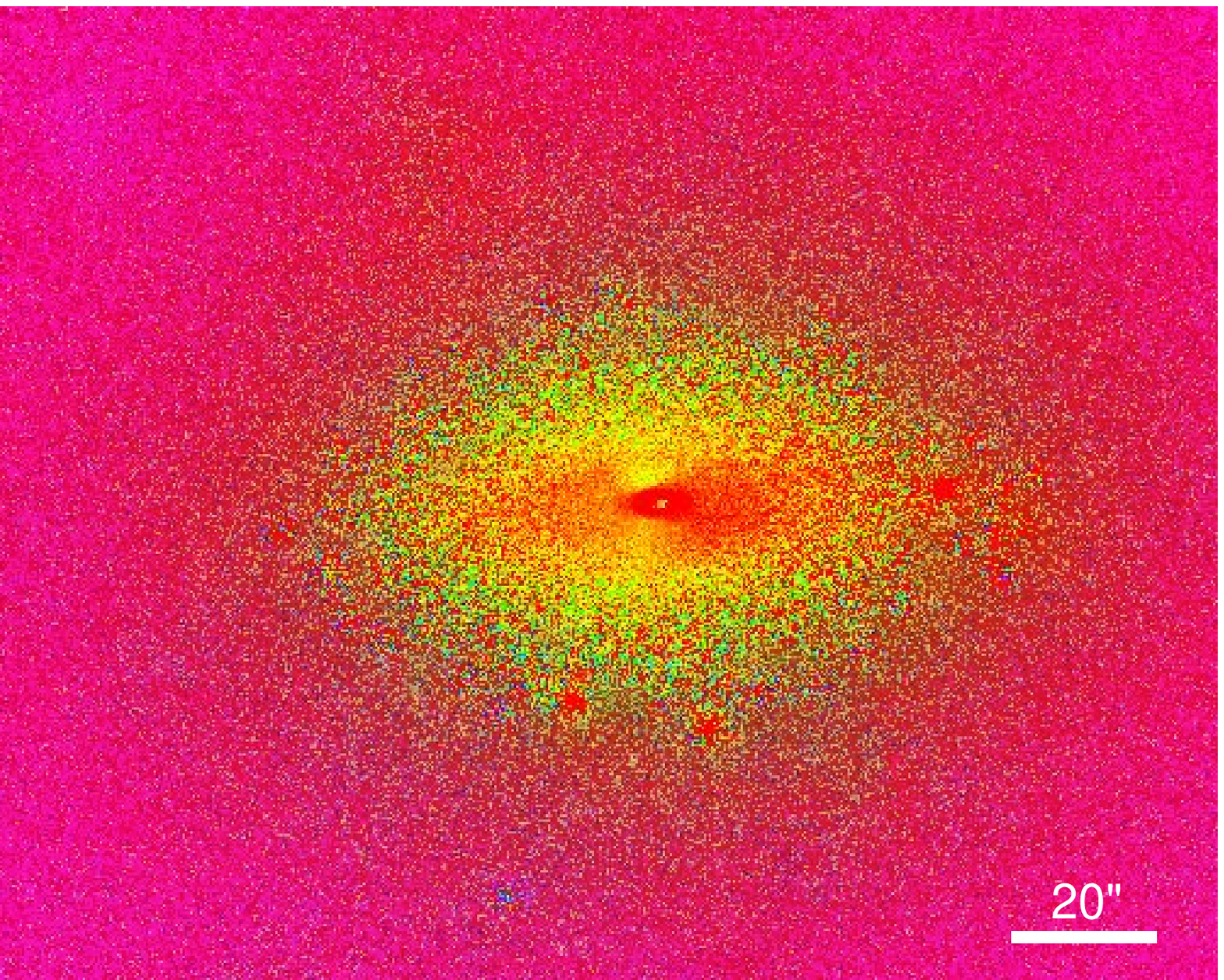}
\plotone{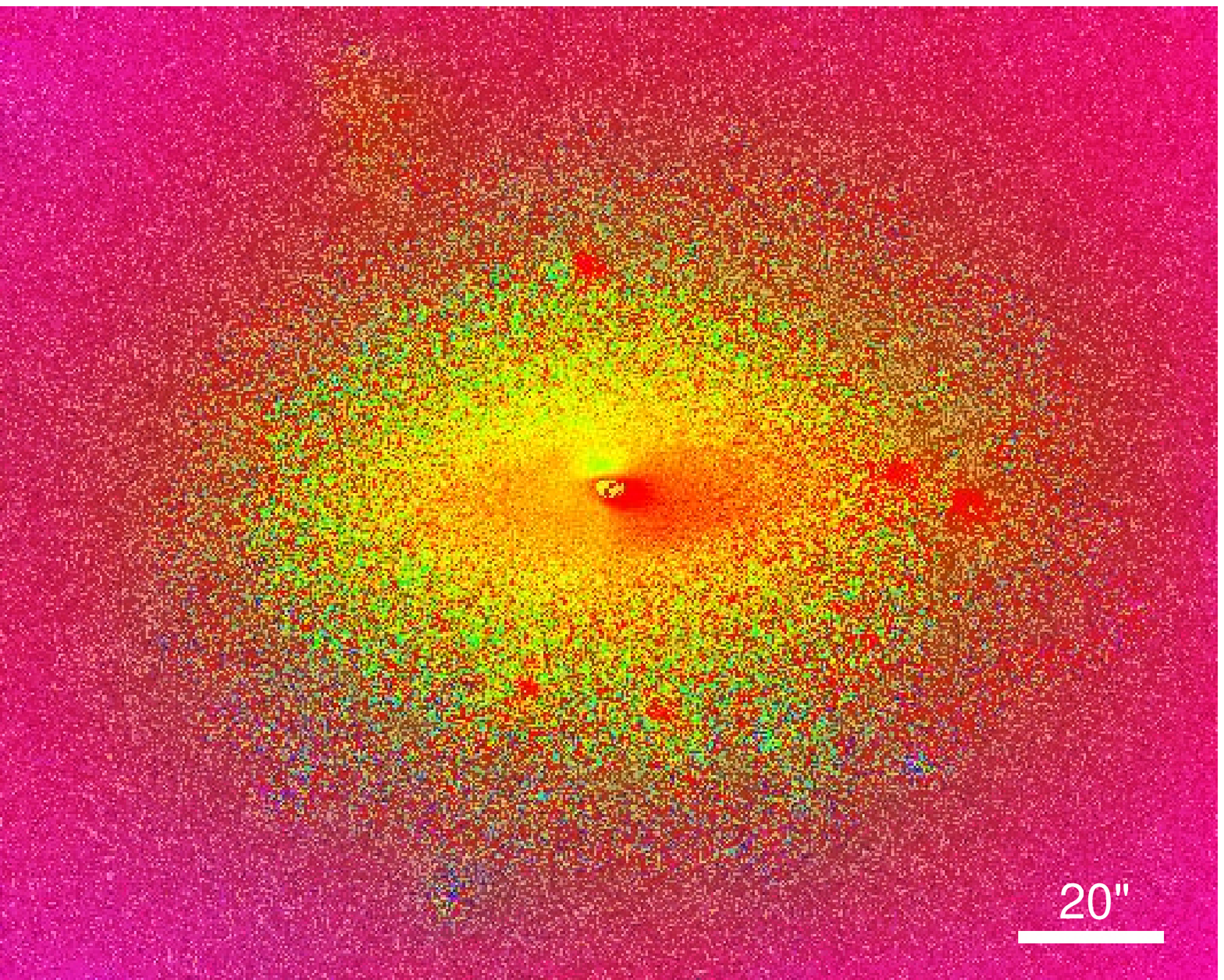}
\caption{a) $B - V$  b)  $B - R $ colour maps of a field containing NGC 4660 from the CCD data. North is to the top and east to the left.   A  `rainbow' colour table is used corresponding to actual colours. }
\end{figure}

\begin{figure}
\epsscale{1}
\plotone{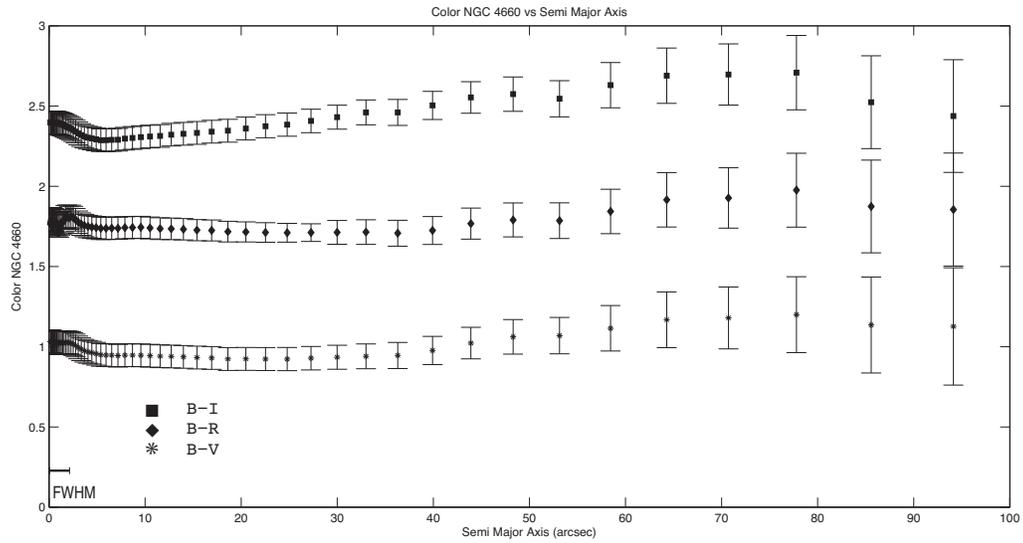}
\caption{$B-V$, $B-R$, $B-I$ radial colour profiles for NGC 4660, from the CCD data. Error bars are included. The colour profiles are mainly flat in the central regions, though $B-I$ rises slightly with radius. $B-V$ is slightly redder at large radius, but the change is $\leq 0.2$ mag.  The seeing FWHM of the data is indicated with a horizontal line.}
\end{figure}

\begin{figure}
\epsscale{1}
\plotone{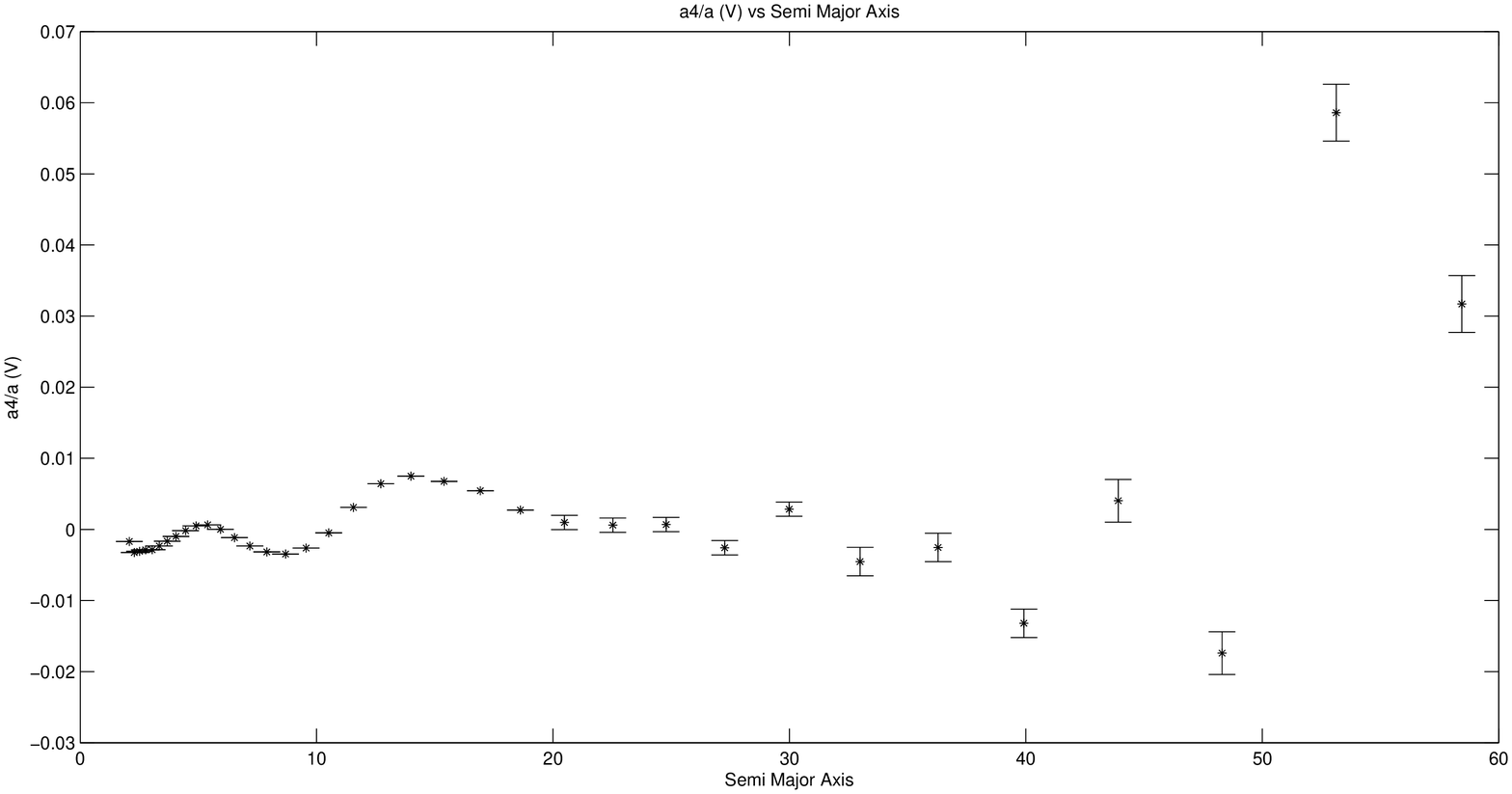}
\plotone{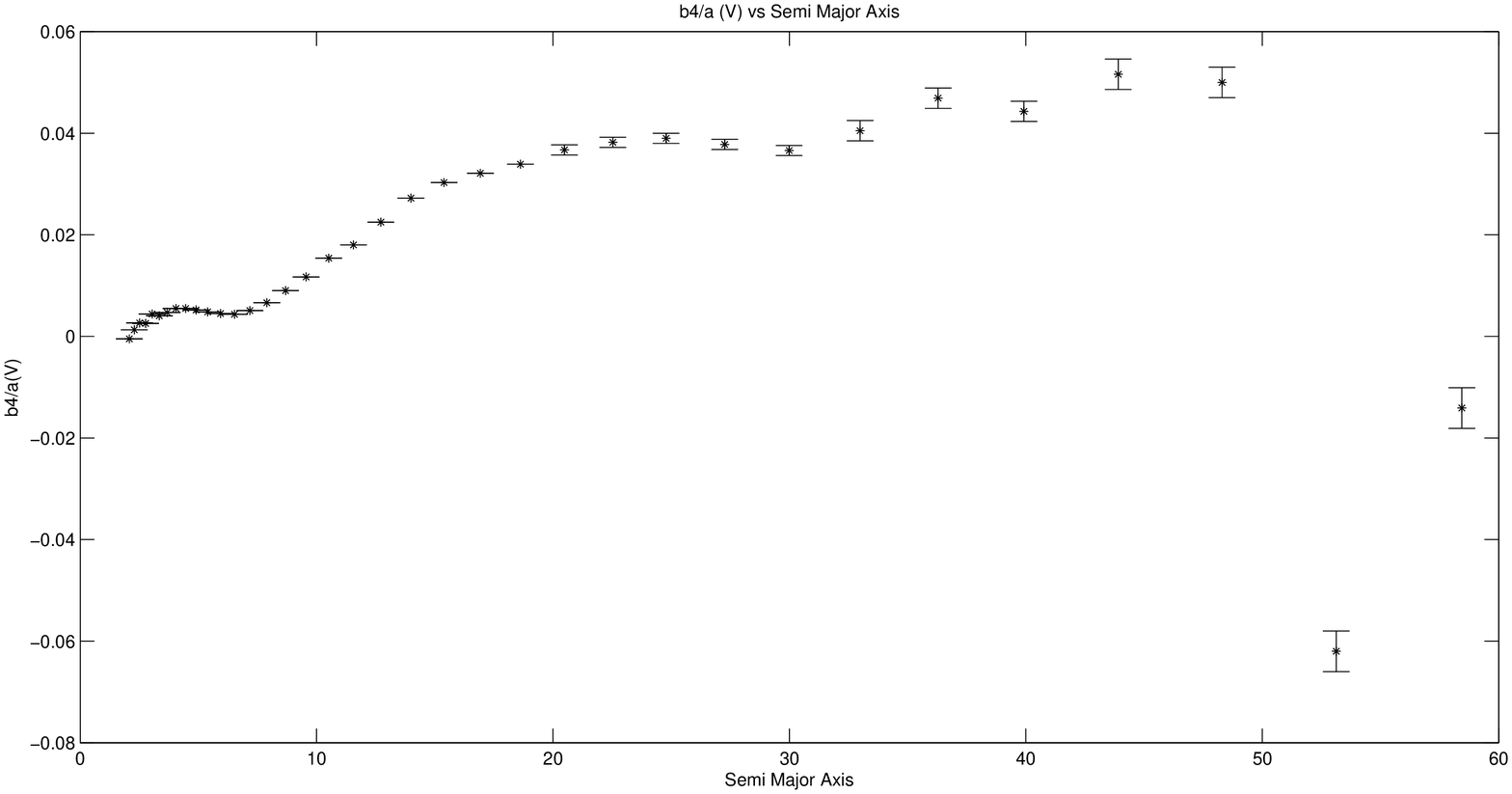}
\caption{a)  $a_4/a$ profile for NGC 4660 in the V band from the CCD data. Positive values $< 0.01$ appear in the region of the disk, while there are large, mainly positive values for $a > 50 $ arcsec  b) $b_4/a$ profile for NGC 4660 in the V band. There are large positive `disky' values for the whole region $a < 45$ arcsec.  Here $b_4/a$ is used as the 4th cosine coefficient. In both plots the innermost data point is at 3 arcsec, outside the seeing FWHM of 2 arcsec. }
\end{figure}

\begin{figure}
\epsscale{1}
\plotone{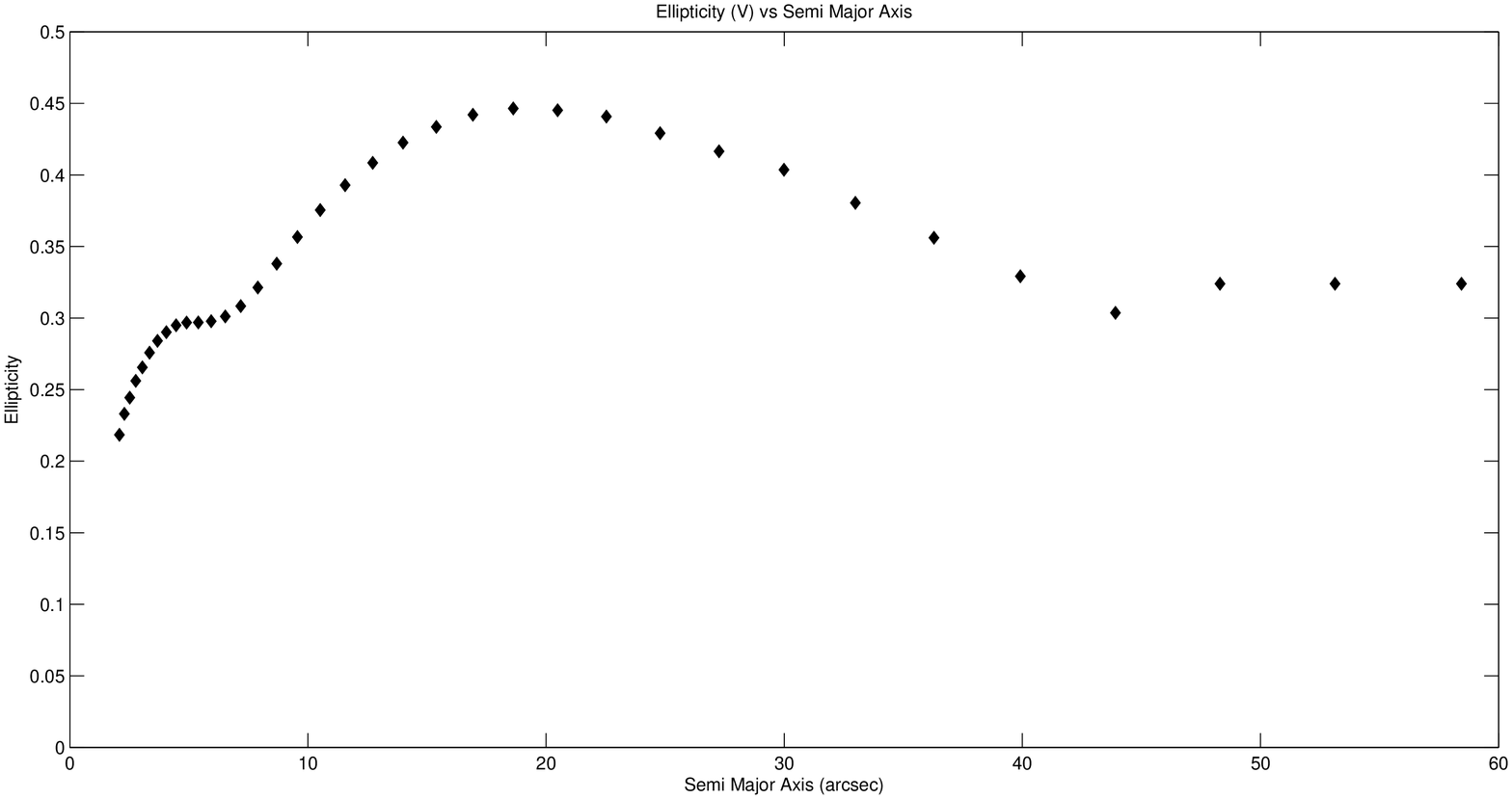}
\plotone{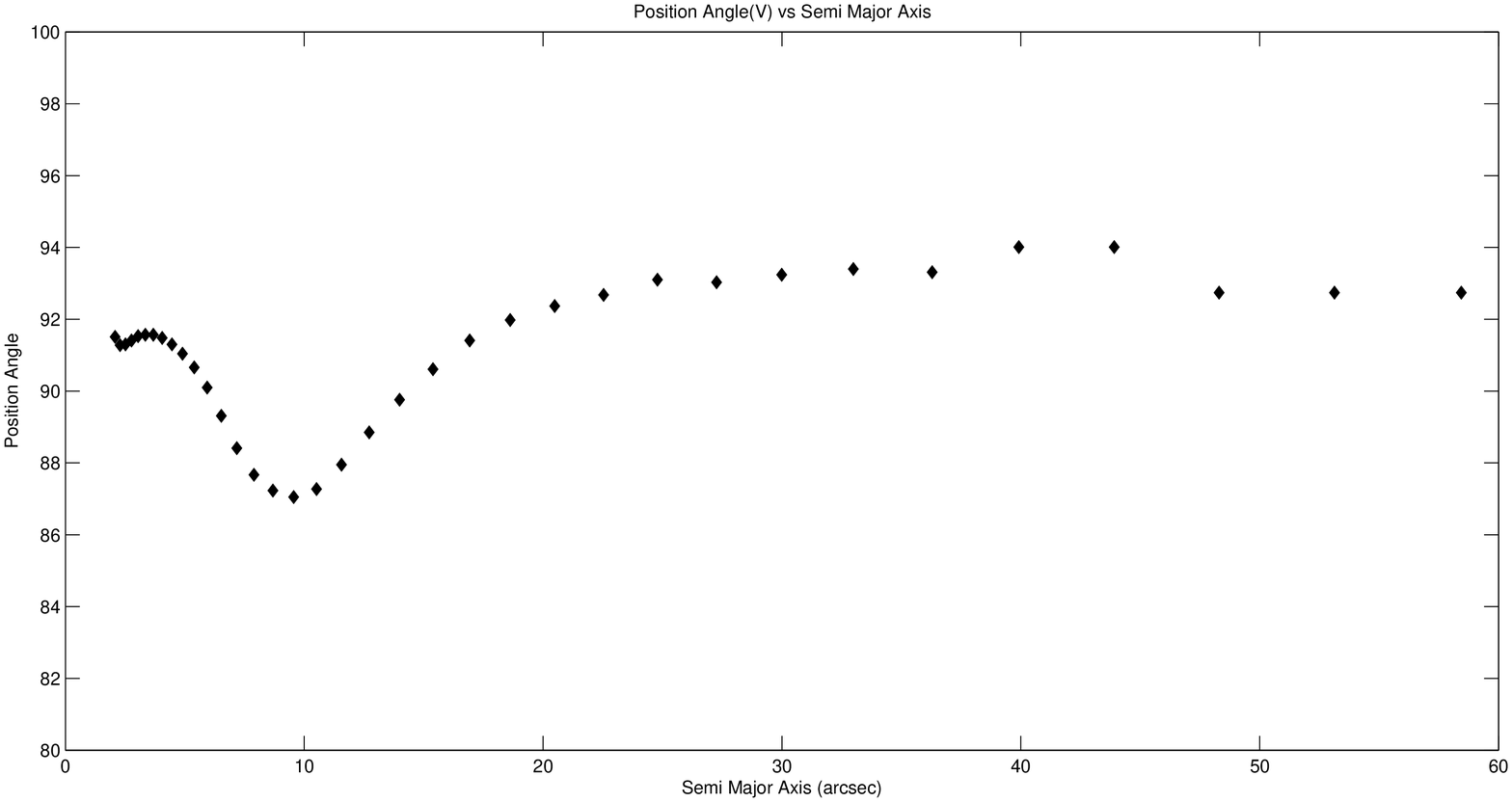}
\caption{a) Ellipticity profile of NGC 4660 in the V band from the CCD data. The maximum in the profile seen around 20 arcsec corresponds to the disk component  b) Position angle profile of NGC 4660 in V. There is a slight twist from 92 to 87 degrees between 3--10 arcsec and again from 87 to 93 degrees between 10--20 arcsec. These correspond to the presence of a bar in the disk component.  In both plots the innermost data point is at 3 arcsec, outside the seeing FWHM of 2 arcsec.}
\end{figure}

\begin{figure}
\epsscale{.65}
\plotone{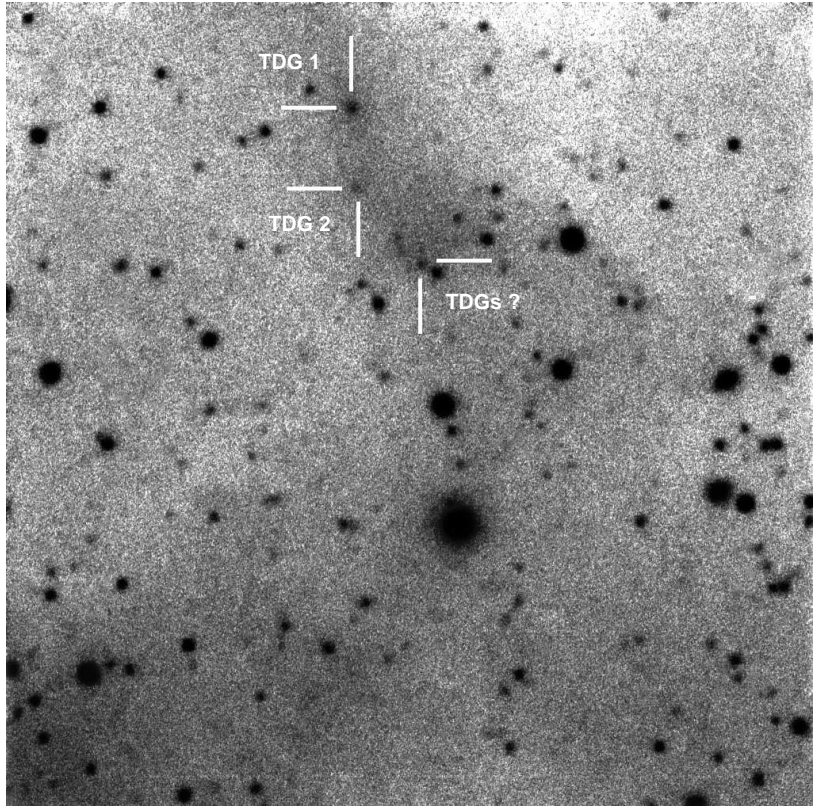}
\caption{CCD field containing the brightest part of the filament of NGC 4660, centred at RA = 12:44:17 Dec. = +11:14 (J2000) with field of view $5.3 \times 5.3$ arcmin$^{2}$. The exposure is 60 min in the R band. The filament is seen as an area of diffuse emission in the north-centre of the field  (running NNE-SSW)
and just S of centre between two stars. There is also an area of emission in the SE of the field corresponding to emission from the NW part of the halo of NGC 4660. The northern detection of the filament  has two faint peaks of surface brightness in the north-centre of the field which may correspond to Tidal Dwarf Galaxies,  marked as TDG 1 and TDG 2, while some other possible candidates are marked as TDGs? }
\end{figure}

\begin{figure}
\epsscale{.95}
\plotone{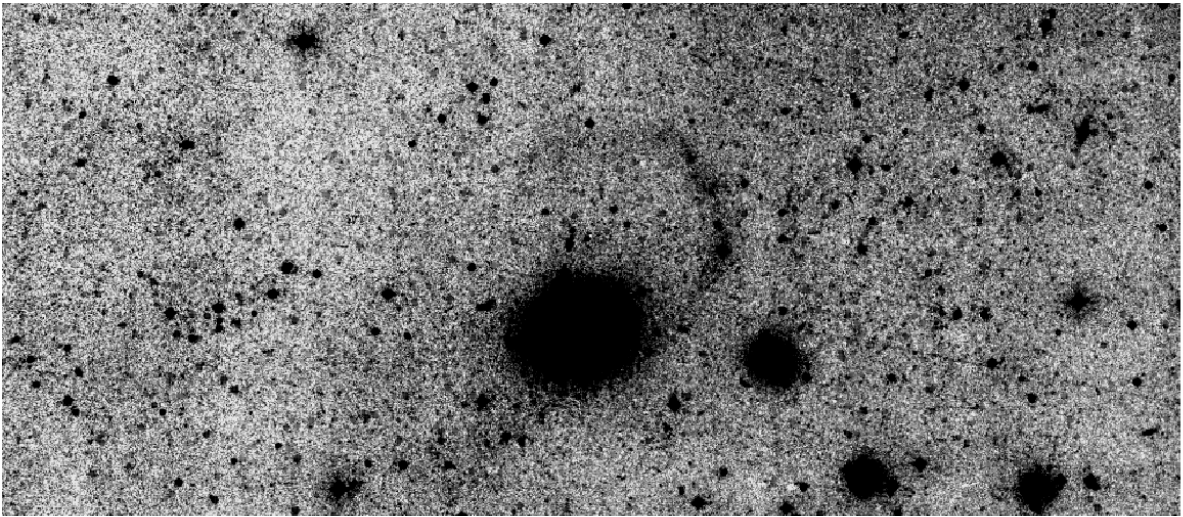}
\caption{Field containing NGC 4660 and its filament, NW of the galaxy, from the co-added Schmidt data. The field is 30 x 15 arcmin$^{2}$. North is to the top and east to the left. }
\end{figure}


\end{document}